\documentclass[11pt,english]{article}

\usepackage{makeidx}  
\usepackage{xspace}
\usepackage{amsmath}
\usepackage{amssymb}
\usepackage{graphicx}
\usepackage[en,nl]{bdalgo}

\newcommand{\Latin}[1]{\textit{#1}}
\newcommand{\eg}{\Latin{e.g.},\xspace}
\newcommand{\ie}{\Latin{i.e.},\xspace}

\newtheorem{proposition}{Proposition}

\newenvironment{proof}[1][0cm]{
  \begin{list}{\bf Proof.~}{
      \setlength{\itemindent}{0cm}
      \setlength{\labelsep}{0cm}
      \setlength{\labelwidth}{#1}
      \setlength{\leftmargin}{#1}
    \item
    }
}{%
  \hfill$\Box$
  \end{list}
}

\usepackage[USletterpaper,simple]{bdpage}
\TextSize{23.8cm}{16.9cm}

\newcommand{\svs}{\vspace{+.5cm}}

\begin{document}

\title{Best-effort Group Service in Dynamic Networks%
       \thanks{Supported by R\'egion Picardie, proj. APREDY.}
}

\author{Bertrand Ducourthial$^\ddagger$ \and Sofiane Khalfallah$^\ddagger$ \and Franck Petit$^\dagger$\\
\begin{small}
 $^\ddagger$\begin{tabular}[t]{l}
   (1) Universit\'e de Technologie de Compi\`egne\\
            (2) CNRS Heudiasyc UMR6599\\
                Centre de Recherche de Royallieu\\
                B.P. 20529, Compi\`egne, France\\
              \end{tabular}\hspace{6mm}%
 $^\dagger$ \begin{tabular}[t]{l}
   (1) Universit\'e PM. Curie\\
            (2) CNRS LIP6 UMR7606\\
            (3) INRIA REGAL\\
	    4, place Jussieu
            75005 Paris, France\\
          \end{tabular}
\end{small}
}

\date{}

\maketitle 

\begin{abstract}
We propose a group membership service for dynamic ad hoc networks.  It maintains
as long as possible the existing groups and ensures that each group diameter is
always smaller than a constant, fixed according to the application using the
groups.
The proposed protocol is self-stabilizing and works in dynamic distributed
systems. Moreover, it ensures a kind of \emph{continuity} in the service offer
to the application while the system is converging, except if too strong topology
changes happen. Such a \emph{best effort} behavior allows applications to rely on
the groups while the stabilization has not been reached, which is very useful in
dynamic ad hoc networks.
\svs

\end{abstract}

\section{Introduction}

\paragraph{Self-stabilization in dynamic networks}
A \emph{dynamic} network can be seen as an (\emph{a priori} infinite) sequence
of networks over time. In this paper, we focus on dynamic \emph{mobile}
networks. Examples of such networks are \emph{Mobile Ad hoc} networks (MANETs) or
\emph{Vehicular Ad hoc} networks (VANETs).

Designing applications on top of such networks require dealing with the lack of
infrastructure~\cite{S02,DBLP:conf/icdcit/JhumkaK07}. 
One idea consists in building virtual structures such as clusters, backbones,
or spanning trees.
However, when the nodes are moving, the maintenance of such structures may
require more control.
The dynamic of the network increases the control overhead.
Thus, distributed algorithms should require less overall organization of the
system in order to remain useful in dynamic networks.

Another paradigm for building distributed protocols in mobile ad hoc networks
consists in designing self-stabilizing algorithms~\cite{BDHY07}. These
algorithms have the ability to recover by themselves (\ie automatically) from an 
inconsistent state caused by transient failures that may affect a memory or a message. 
In this context, the topology changes can be considered as transient failures because 
they lead to an inconsistency in some memories. Indeed, when a node appears or disappears in the
network, all its neighbors should update their neighborhood knowledge.

Self-stabilizing algorithms have been intensively studied the two last decades
for their ability to tolerate transient faults~\cite{D00}.
However, it is important to notice that such algorithms do not ensure all the time
the desirable behavior of the distributed system,
especially when faults occur and during a certain period of time following them. 
In dynamic systems, it becomes illusory to expect an application that continuously 
ensures the service for which it has been designed.
In other words, what we can only expect from the distributed algorithms is to
behave as ``the best'' as possible, the result depending on the dynamic of the
network.

In this paper, we propose a new approach in the design of distributed solutions
for dynamic environments.  
We borrow the term ``{\em best-effort}'' from the networking community to qualify
the algorithms resulting of our approach. 
Roughly speaking, a best-effort algorithm is a self-stabilizing algorithm 
that also maintains an extra property, called {\em continuity}, 
conditioned by the topology changes.  

Continuity aims to improve the output of the distributed protocol during the
convergence phase of the algorithm, provided that a topological property is
preserved.
This means that there is a progression in the successive outputs of the
distributed protocols, except if the network dynamic is too high. This is
important in a distributed system where the
dynamic (that is, the frequent topology changes) can prevent the system to 
converge to the desirable behavior.
Since the output of the protocol will certainly be used before the
stabilization, the continuity ensures that third party applications can rely on
it instead of waiting. The output will certainly be modified in the future, but
without challenging previous ones.

In some aspects, our approach is close to the ones introduced in~\cite{KM06c}
and in \cite{PODC95}.  In~\cite{KM06c}, the authors introduce the notion of
\emph{safe-convergence} which guarantees that the system quickly converges to a
safe configuration, and then, it gracefully moves to an optimal configuration
without breaking safety.  However, the solution in~\cite{KM06c} works on a
static network.  In~\cite{PODC95}, the authors use the notion of \emph{passage
  predicate} to define a {\em superstabilizing} system, \ie a system which is
stabilizing and when it is started from a legitimate state and a single topology
change occurs, the passage predicate holds and continues to hold until the
protocol reaches a legitimate state.
By contrast, the continuity property is intended to be satisfied {\em before} a legitimate
configuration has been reached. It must be satisfied during the stabilization
phase, and between two consecutive stabilization phases (convergence phase
followed by stability phase). 

We illustrate our approach by specifying a new problem, called \emph{Dynamic
  Group Service} inspired from vehicular ad hoc networks (VANET), an emblematic
case of dynamic ad hoc networks. We then design a best effort distributed
protocol called \texttt{GRP} for solving this problem: we prove that it is
self-stabilizing and fulfills a continuity property, allowing applications to
use the groups while the convergence may be delayed because of the dynamic of
the network.

\paragraph{Dynamic group service}
Vehicular ad hoc networks currently attract a lot of attention \cite{BEH04}.
Many VANET applications require cooperation among close vehicles during a given
period: collaborative driving, distributed perception, chats and other
infotainment applications.
Vehicles that collaborate form a \emph{group}. A group is intended to grow
until a limit depending on the application. For instance, the distributed
perception should not involve too far vehicles, a chat should be responsive
enough, that limits the number of hops, etc.
When the group diameter is larger than the bound given by the application, it
should be split into several smaller groups.
However, a group should not be split if this is not mandatory by the diameter
constraint in order to ensure the best duration of service to the application
relying on it.
Even if another partitioning of the network would have been better (\eg less
groups, no isolated vehicle), it is preferable to maintain the composition of
existing groups. It is expected that, thanks to the mobility of the nodes, small
groups will eventually succeed in merging. It is then more important to 
maintain existing groups as long as possible.

\paragraph{Best-effort \texttt{GRP} algorithm}
To solve the \emph{Dynamic Group Service}, we propose a best-effort distributed
algorithm called \texttt{GRP} (for \texttt{GRouP}) designed for unreliable
message passing systems.
This algorithm stabilizes the views (the local knowledge of the group to which
belongs the node) in such a way that all the members of a group will eventually
share the same view (in which only the members appear). The groups' diameters
are smaller than a fixed applicative constant \texttt{Dmax} and neighbor groups
merge while the diameter constraint is fulfilled. Moreover, our algorithm admits
the following continuity property: no node disappears from a group except if a
topology change leads to the violation of the diameter constraint.
This allows to the applications requiring the groups (\eg chat) to run before
the convergence of \texttt{GRP}, that may be delayed because of the dynamic of
the network.

To the best of our knowledge, only a few number of papers address the problem
of group membership maintenance in the context of self-stabilization.  Recently,
in~\cite{DLV08}, the authors propose a self-stabilizing $k$-clustering algorithm
for static networks.  In~\cite{TMC06}, the authors propose a self-stabilizing
group communication protocol.  It relies on a mobile agent that collects and
distributes information during a random walk.  This protocol does not allow 
building groups that strech over at most $k$ hops.

Group communication structures have been proposed in the literature to achieve
fault-tolerance in distributed systems~\cite{ACM93}, by providing for instance
replication, virtual synchrony, reliable broadcast, or atomic broadcast
(\emph{e.g.},~\cite{S90,GS97}).
Other works deal with the $k$-clustering or $k$-dominating set problem, \eg
\cite{APHV00,CKV01,Demirbas06,Johnen09,KM06c,KP98,PB04}, where nodes in a group are at
most at distance $k$ from a \emph{cluster-head} or \emph{dominant node}. The aim
of these algorithms is to optimize the partitioning of the network.
The group service we propose in this paper is different in the sense that its aim is neither to
optimize any partitioning nor to build group centered to some nodes. Instead, it
tries to maintain existing groups as long as possible while satisfying a
constraint on the diameter, without relying on a specific node (that may move or
leave).

\paragraph{Organization}
In Section~\ref{s:model}, we describe the distributed system we consider in this
paper. We also state what it means for a protocol to be self-stabilizing and
best effort regarding a continuity property conditioned by topology changes.
Next, in Section~\ref{s:spec}, we specify the Dynamic Group Service problem and
in Section~\ref{s:algo}, we describe our \texttt{GRP} algorithm solving it%
\footnote{\label{refonline}.  Note that the algorithm has been successfully implemented using the
  Airplug software suite. The detailed algorithm used for the implementation is
  available on our website (as long as the software):\newline \noindent
{\tt http://www.hds.utc.fr/$\sim$ducourth/airplug/doku.php?id=en:dwl:grp:accueil}.
Some screenshot movies are also available here:\newline \noindent
{\tt http://www.hds.utc.fr/$\sim$ducourth/airplug/doku.php?id=en:doc}
}.
The proofs are given in Section~\ref{s:proof}. Finally, we make some
concluding remarks in Section~\ref{s:conclu}.
By lack of place, some proofs are in appendix.

\section{Model}\label{s:model}

We define the distributed system $\mathcal{S}$ as follows.

\paragraph{System}
Let $V$ be the set of nodes, spread out in an Euclidean space.  The total number
of nodes in $V$ is finite but unknown.  Each node is equipped with a processor
unit (including local memory) and a communication device. A node can move in the
Euclidean space.  It is either \emph{active} or \emph{inactive}. When it is active, it can
compute, send and receive messages by executing a local algorithm.  The
\emph{distributed protocol} $\mathcal{P}$ is composed of all the local
algorithms.

We define the \emph{vicinity} of a node $v$ as the part of the Euclidean space
from where a node $u$ can send a message that can be received by $v$ (the
vicinity depends on the communication devices, the obstacles, etc.).  A node $v$
can receive a message from $u$ if ($i$) both $u$ and $v$ are active, ($ii$) $u$
is in the vicinity of $v$, ($iii$) $u$ is sending a message, ($iv$) no other
node in the vicinity of $v$ is currently sending a message, and ($v$) $v$ is not
sending a message itself (any active node that is not sending is able to
receive).

We assume that on each node the message sending is driven by a {\em timer}.
We admit the following \emph{fair channel} hypothesis: there exists two time
constants $\tau_1$ and $\tau_2$ with $\tau_1 \geq \tau_2$ such that, starting from
a date $t$, any node $v$ is able to receive before the date $t+\tau_1$ a message
from each node $u$, providing that $u$ is in the vicinity of $v$ between $t$ and
$t + \tau_1$ and attempts to send a message every $\tau_2$ units of time.
At any time instant $t$, there is a \emph{communication link} from $u$ to $v$ if
both $u$ and $v$ have the state active (at $t$), and if $u$ is into the vicinity
of $v$ (at $t$).  A communication link is oriented because $u$ could be in the
vicinity of $v$ while the converse is false.

We assume the following hypotheses, close to the IEEE~802.11 protocol.  
($i$) The communication channel contains at most one message---{\em one-message channel}.
($ii$) If a node $u$ keeps continuously sending a message $m$, then $u$ 
eventually sends $m$---{\em fair sending}. 
($iii$) If a node $u$ keeps continuously receiving a message, then $u$ 
eventually receives a message---{\em fair reception}. 
($iv$) If a node $u$ has continuously an action $a$ to execute, then $u$ executes
$a$ in finite time---{\em fair activation}.

A \emph{configuration} $c$ of $\mathcal{S}$ is the union of states of memories
of all the processors and the contents of all the communication links. An empty
communication link is denoted in the configuration by a link that contains an
empty set of messages. By the way, there is a single topology per configuration.
Let $\mathcal{C}$ be the set of configurations.
An \emph{execution} of a distributed protocol $\mathcal{P}$ over $\mathcal{S}$
is a sequence of configurations $c_0, c_1, \ldots$ of $\mathcal{S}$ so that
$\forall i \leq 0$, $c_i$ moves to $c_{i+1}$ by changing the memory of at least one process, including its message
buffers (\ie by sending or receiving messages).  

We denote by $G^{c_i}$ the topology of $\mathcal{S}$ during the configuration
$c_i$.
In a \emph{static} system $\mathcal{S}$, we have $G^{c_i}=G^{c_0}$ in every
execution $c_0,c_1,c_2,\ldots$.  Otherwise, the system $\mathcal{S}$ is said to
be \emph{dynamic}.

\paragraph{Self-Stabilization}
Let $\mathcal{X}$ be a set. Then $x\vdash \Pi$ means that an element $x\in
\mathcal{X}$ satisfies the predicate $\Pi$ defined on the set $\mathcal{X}$ and
$X \vdash \Pi$ with $X \subset \mathcal{X}$ means that any $x \in X$ satisfies
$x\vdash \Pi$.
We define a special predicate $\mathsf{true}$ as follows:
$\forall x\in \mathcal{X}$, $x\vdash \mathsf{true}$.
Let $\Pi_1$ and $\Pi_2$ be two predicates defined on the set of configurations
$\mathcal{C}$ of the system $\mathcal{S}$.  $\Pi_2$ is an {\em attractor} for
$\Pi_1$ if and only if the following condition is true: for any configuration
$c_1 \vdash \Pi_1$ and for any execution $e = c_1,c_2, \ldots $, there exists $i
\geq 1$ such that for any $j \geq i$, $c_j \vdash \Pi_2$.

Define a \emph{specification} of a task as the predicate $\Pi$ on the set
$\mathcal{C}$ of configurations of system $\mathcal{S}$.
A protocol $\mathcal{P}$ is self-stabilizing
for $\Pi$ if and only if
there exists a predicate $\mathcal{L_P}$ (called the legitimacy predicate)
defined on $\mathcal{C}$ such that the following conditions hold:\newline
\noindent 1. For any configuration $c_1 \vdash \mathcal{L_P}$, and for any execution $e =
c_1,c_2, \ldots$, we have $e \vdash \Pi$ (correctness).
\newline
\noindent 2. $\Pi$ is an attractor for \textsf{true} (closure and convergence).

\paragraph{Best effort requirement}
We denote by $\Pi_T$ a \emph{topological} predicate defined on the pairs of
successive configurations in an execution. Such a predicate is intended to be
false when an ``important topology'' change happens.
We denote by $\Pi_C$ a \emph{continuity} predicate defined on the pairs of
successive configurations in an execution. Such a predicate is intended to be
false when the quality of the outputs produced by protocol
$\mathcal{P}$ in the two successive configurations decreases.

The protocol $\mathcal{P}$ offers a best effort continuity of services if $\Pi_T
\Rightarrow \Pi_C$.

\section{Dynamic Group Service Problem}\label{s:spec}

The Dynamic Group Service protocol is inspired from applications requirements in
Vehicular Ad Hoc networks (VANET), such as collaborative perception or
infotainment applications.

\paragraph{Informal specification} On each node $v$, a variable
$\texttt{view}_v$ gives the composition of the group to which $v$ belongs. This
will be used by the applications. The agreement property says that all nodes in
group of $v$ agree on the composition of the group. The safety property says
that the diameter of each group is smaller than a constant $\texttt{Dmax}$. The
maximality property says that small groups merge to form larger groups.

To deal with the dynamic of the network, the algorithm should be able to satisfy
these three properties in finite time after the last failure or topology change
(self-stabilization). To allow the applications to run while the convergence has
not been reached, the algorithm should ensure a best effort requirement: if the
distance between the members of a group remains smaller than $\texttt{Dmax}$
(topological property), then no node will leave the group (continuity
property). This is important because the convergence may be delayed because of
the dynamic of the network.

\paragraph{Formal specification}
Let $G(V,E)$ be a graph.  Let $d(u,v)$ be the \emph{distance} between $u$ and
$v$ (length of the shortest path from $u$ to $v$ in $G$).
A \emph{subgraph} $H(V_H,E_H)$ is defined as follows: $V_H \subseteq V$ and
$\forall (u,v) \in E,\ (u \in V_H \mbox{ and } v \in V_H) \Rightarrow (u,v) \in
E_H$.
Two subgraphs $H_1(V_1,E_1)$ and $H_2(V_2,E_2)$ of a graph $G$ are said
\emph{distinct} if $V_1 \cap V_2 = \emptyset$.
Let $X \subseteq V$ be a set of nodes. We denote by $d_X(u,v)$ the distance
between $u$ and $v$ in the subgraph $H(X,E_H)$, that is, the length of the
shortest path from $u$ to $v$ with only edges of $E_H$. If such a path does not
exists, then $d_X(u,v) = +\infty$.

Given a graph $G$, the problem considered in this paper consists in designing a
distributed protocol that provides a partition of $G$ into disjoint subgraphs
called {\em groups} that satisfies constraints described
below.  
Denote by $\texttt{view}_v^c$ the knowledge of $v$ about its group in
configuration $c$ (output on node $v$).

Let $\Pi_A$ be the predicate defined on the configurations and called
\emph{agreement property}: $\Pi_A(c)$ holds if and only if there exists a
partition of disjoint subgraphs $H_1(V_1,E_1)$, $H_2(V_2,E_2)$ , $\ldots,$
$H_i(V_i,E_i), \ldots$ of $G(V,E)$ such that for every nodes $u,v \in V$, $(u \in V_i
\mbox{ and } v \in V_i) \Leftrightarrow \texttt{view}_u^c = \texttt{view}_v^c =
V_i$.

Let $\Omega_v^c$ be the \emph{group} of $v$ in configuration $c$, defined by:
($i$) $\Omega_v^c = \texttt{view}_v^c$ if $v \in \texttt{view}_v^c$ and $\forall u
\in \texttt{view}_v^c, \texttt{view}_v^c = \texttt{view}_u^c$, ($ii$) $\Omega_v^c
= \{v\}$ otherwise.  Note that given any configuration $c$, if $\Pi_A(c)$ holds,
$\{ \Omega_v^c, v \in V\}$ defines a partition of $G$ into disjoint subgraphs of
$G$, {\it i.e.}, there exists a partition of disjoint subgraphs $H_1(V_1,E_1)$,
$H_2(V_2,E_2)$, $\ldots$, $H_i(V_i,E_i), \ldots$ such that $\forall v \in V_i,\
\Omega_v^c = V_i$ for every subgraph $H_i$. 

Let $\texttt{Dmax}$ be an integer representing the maximal admissible distance between two nodes belonging to
the same group.  
Let $\Pi_S$ be the predicate defined on the configurations and called \emph{safety
property}: $\Pi_S(c)$ holds if each group is connected and its diameter is
smaller than $\texttt{Dmax}$.
More formally $\Pi_S(c) \equiv \forall v \in V$, $\max_{x,y \in \Omega_v^c}
d_{\Omega_v^c}(x,y) \leq \texttt{Dmax}$.

Let $\Pi_M$ be the predicate defined on the configurations and called
\emph{maximality property}: $\Pi_M(c)$ holds if by merging two existing groups,
we cannot obtain a partition satisfying the safety property. More formally
$\Pi_M(c) \equiv \forall u,v \in V$ with $\Omega_u^c \neq \Omega_v^c$, $\exists
x,y \in \Omega_u^c \cup \Omega_v^c$, $d_{\Omega_u^c \cup \Omega_v^c}(x,y) >
\texttt{Dmax}$.

The problem considered in this paper is to design a self-stabilizing protocol
regarding predicates $\Pi_A \wedge \Pi_S \wedge \Pi_M$: after the last failure
or topology change, the algorithm converges in finite time to a behavior where
$\Pi_A$, $\Pi_S$, and $\Pi_M$ are fulfilled.

Note that the above requirement is suitable for fixed topologies only.  The
following predicate deals with dynamic system, {\it i.e.}, with topological
changes.
Let $G^c(V^c,E^c)$ be the graph modeling the topology of the system at
configuration $c$. We introduce the following notation: $d^c$ refers to the
distance in the graph $G^c$, and $d_X^c(u,v)$ denotes the
distance between $u$ and $v$ in $G^{c}$ by considering only edges of the
subgraph $H(X,E_H)$ of $G^c$.
Define the {\em topological property} as the predicate $\Pi_T$ defined on any
couple of two successive configurations $c_i, c_{i+1}$ of an execution $e$ as
follows: $\Pi_T(c_i, c_{i+1})$ holds if, for any pair of nodes belonging to the
same group in $c_i$, the distance between them will still be smaller than
\texttt{Dmax} in $c_{i+1}$. In other words, if a topology change occurred
between $c_i$ and $c_{i+1}$, it has preserved the maximal distance condition.
More formally, $\Pi_T(c_i, c_{i+1}) \equiv \forall v \in V$, $\max_{x,y \in
  \Omega_v^{c_i}} d_{\Omega_v^{c_i}}^{c_{i+1}}(x,y) \leq \texttt{Dmax}$.

Finally, we are looking for protocols attempting to preserve a group partition when a topology change occurs. 
Let $\Pi_C$ be the predicate defined on the couples of successive configurations
and called \emph{continuity property}: $\Pi_C(c_i,c_{i+1})$ holds if in any
group, no node disappears. In other words, an application can work with the
given view because it defines a group in which no node will disappear. More
formally, $\Pi_C(c_i,c_{i+1}) \equiv \forall v \in V$, $\Omega_v^{c_i}
\subset \Omega_v^{c_{i+1}}$.
Obviously, if the dynamic of the network is too large, such a property cannot be
satisfied.  We then introduce the best effort requirement: $\Pi_T \Rightarrow \Pi_C$.

\section{\texttt{GRP} distributed protocol}\label{s:algo}

The GRP distributed protocol is designed for solving the Distributed Group
Service problem in an unreliable message passing system.

\subsection{Principle of the \texttt{GRP} distributed protocol} 
For each node $v$, the candidates to form a group are neighbors up to distance
$\texttt{Dmax}$.  Each node $v$ periodically echanges messages with its neighbors and 
maintains a list of nodes being at distance at most $\texttt{Dmax}$. 
Each sent message sent by $v$ contains the list of $v$.  The list of $v$ contains nodes at 
distance at most $\texttt{Dmax}$ that are in the group or candidates to join the group. 

Our mechenism needs to take in account symmetric links only, \ie links between pairs of nodes
$u$ and $v$ so that if $v$ is considered by $u$ as a neighbor, then $u$ 
(resp. $v$) is considered as a neighbor by $v$. 
In order to implement this, we use marks.  Each node proceeds as follows:
if $v$ receives a
list from $u$ that does not contain itself, then it adds $\underline{u}$ in its
list (which will be sent to the neighbors at the next timer expiration). To the
converse, when $v$ receives a list from $u$ that contains either $v$ or
$\underline{v}$, then it adds $u$ in its list. Marked nodes are not propagated
farther than the neighborhood.

Malformed lists are rejected (such as lists larger than $\texttt{Dmax}$).
Moreover, when a node $v$ receives a list from $u$ which is too long compared to
its current list, it rejects it to avoid any split of its current group. In this
case, $v$ adds $\underline{\underline{u}}$ in its list, meaning that $u$ and $v$
cannot belong to the same group. To the converse, if the received list is not
too long, it is merged with the current list, meaning that $u$ enters to the
group of $v$. Symmetrically, $u$ will accept $v$ in its group.

Several nodes may be accepted concurrently by distant members of a given
group. In some cases, a too large group may be obtained. Then one of the new
members must leave the group (instead of splitting the existing group).  To
avoid any inopportune change in the views (which are used by the applications),
a new member enters in the view of a node only after the end of its
\emph{quarantine} period. This allows guaranteeing that its arrival has been
approved by all the members (no conflicts). A node arrival is propagated to the
group's members in $O(\texttt{Dmax})$; this defines the quarantine period
duration.

When it is necessary to chose which node has to leave the group (to fulfill the
diameter constraint), the choice is done using a \emph{priority} computed by Function~$texttt{pr}$. 
Priorities are
totally ordered; if $\texttt{pr}(u) < \texttt{pr}(v)$, then $u$ has the
priority. 
A powerful implementation of priorities is the oldness of nodes in the groups:
the priority of a node is incremented  by a logical clock~\cite{L78}, except if it belongs to a
group (of more than one node) in which case the priority remains stable. The
last entered nodes in a group have then less priority than the nodes entered
before them.

Priorities on the nodes allow to easily define priorities on the groups by
taking the smallest priority of the members.  Priorities on the groups allows to
ensure the merging of neighbor groups (and the maximality property $\Pi_M$) in
particular cases (loop of groups willing to merge).

\subsection{Building the lists}\label{s:ant}
In the sequel, a node $v$ is an {\em ancestor} of node $u$ if a path exists from $u$ to $v$. 
The messages sent to the neighbors contain \emph{ordered list of ancestors'
  sets}.  The \emph{ordered list of ancestors' sets} of a node $v$ is defined
by:
$\left(a_v^0, \right.$ $a_v^1,$ $\ldots,$ $\left. a_v^{p}\right)$ where any node
$x \in a_v^i$ satisfies $d(x,v) = i$ ($a_v^0 = \{ v\}$) and $p$ is the distance
of the farthest ancestor of $v$.

Computations are done using the $r$-operator $\mathrm{ant}$
\cite{JACIC06,DT03,SSS07}.
Let $\mathbb{S}$ be the set of lists of vertices' sets. For instance, if
$a,b,c,d,e$ are vertices, $\left( \{d\},\{b\},\{a, c\}\right)$ and $\left(
\{c\},\{a,e\},\{ b\}\right)$ belong to $\mathbb{S}$.
Let $\oplus$ be the operator defined on $\mathbb{S}$ that merges two lists while
deleting needless or repetitive information (a node appears only one time in a
list of ancestors' sets).  For instance: \newline
\noindent
$ \left( \{d\},\{b\},\{a, c\}\right) \oplus \left( \{c\},\{a,e\},\{ b\}\right) = $
\newline \noindent
$\left( \{d ,c\}, \{b,a,e\}, \{a,c,b\} \right) = \left(\{d,c\}, \{b,a,e\}\right)$.

Finally, let $r$ be the endomorphism of $\mathbb{S}$ that inserts an empty set at
the beginning of a list. For instance: \newline \noindent
$r( \{d\},\{b\},\{a,c\} )=(\emptyset, \{d\}, \{b\}, \{a,c\})$.

We then define the operator $\mathrm{ant}$ by: $\mathrm{ant}(l_1, l_2) = l_1
\oplus r(l_2)$, where $l_1$ and $l_2$ are lists belonging to $\mathbb{S}$. This
is a strictly idempotent $r$-operator \cite{SSS07} inducing a partial order
relation. It leads to self-stabilizing static tasks (building the complete
ordered lists of ancestor sets) in the register model \cite{DT03}.
Since our wireless communication model admits bounded links, these results can
be extended to this model. (Refer to the discussion related to $r$-operators in
wireless networks in \cite{JACIC06}.)

\subsection{\texttt{GRP} algorithm}

Each node $v$ computes its output (\texttt{list}$_v$, \texttt{view}$_v$ and the
priorities) when its timer $T_c$ expires. It broadcasts its output in the
neighborhood when the timer $T_s$ expires ($T_s \leq T_c$).
All messages received from the neighborhood are collected on $v$ in
\texttt{msgSet}$_v$. If a neighbor sends more than one message before the timer
expiration, only the last received is kept.
After computation, the variable \texttt{msgSet}$_v$ is reset in order to detect
when a neighbor leaves.

\begin{footnotesize}
\begin{algorithm*}{\texttt{GRP}, node \texttt{v}}
\ACTION{Upon reception of a message \texttt{msg} sent by a node \texttt{u}}
  \I update message of $u$ in \texttt{msgSet}$_v$
\ENDACTION
\ACTION{Upon $T_c$ timer expiration}
  \I compute()
  \I reset \texttt{msgSet}$_v$
  \I restart timer $T_c$ with duration $\tau_1$
\ENDACTION
\ACTION{Upon $T_s$ timer expiration}
  \SEND{$\texttt{list}_v$ with priorities} to the neighbors \label{l:send}
  \I restart timer $T_s$ with duration $\tau_2$
\end{algorithm*}
\end{footnotesize}

A computation (in procedure \texttt{compute}, below) consists in building the
ordered list of ancestor' sets as well as the view. The list is sent to the
neighbors to be used in their $\mathrm{ant}$ computation. The view is the output
of the protocol used by the applications (\eg chat, collaborative perception...)
which requested the \texttt{GRP} algorithm, and which determined the diameter
constraint \texttt{Dmax} (fixed during all the execution).

First, the incoming lists are checked.  Line~\ref{l:good}, when the list sent by
$u$ and received by $v$ does not contain $v$, is malformed or is too
long\footnote{$s(\texttt{list})$ returns the number of elements in
\texttt{list}; $\texttt{list}.i$ returns the $i$th element of \texttt{list},
starting from 0.}, it is replaced by $(\underline{u})$.  When $u$ receives the
list of $v$ containing $\underline{u}$, it accepts the list of $v$ and sends a
list containing $v$. Thanks to this triple handshake, the link has been detected
as symmetric (by the way, asymmetric link information are not propagated).

Line~\ref{l:cond}, if the received list is too long, the sender $u$ is marked as
incompatible ($\underline{\underline{u}}$).  Roughly speaking, a list received
by a node $u$ from another node $v$ is compatible if, by combining its list with
the one of $v$, $u$ does not increase the diameter of its group beyond
\texttt{Dmax}.
In order to reach this goal, it is enough to test if the sum of the lengths of
both lists is less than or equal to $\texttt{Dmax}+1$.  But, such simple test
would avoid merging two groups by taking advantage of short cuts between both
groups.  In other words, this would ignore the knowledge that nodes of a group
have on nodes belonging to the other group.  The technical condition used in
Function~\texttt{compatibleList()} deals with such an optimization.

Then a first computation is performed using the $\mathrm{ant}$ operator. Thanks
to the \texttt{goodList} test, the sizes of the incoming lists are smaller than
$\texttt{Dmax}+1$.  However, the computed list could reach the size of
$\texttt{Dmax} + 2$ while the maximum is $\texttt{Dmax} + 1$ (the $\mathrm{ant}$
operation increases by one the list sizes).
In this case, a choice has to be done between either the local node $v$ or the
farthest nodes in the received lists. This choice is done by using priorities,
Line~\ref{l:order}. If nodes belong to the same group, node priorities are
compared. If nodes are not in the same group, this is a group merging and group
priorities are compared (to avoid loops of groups willing to merge).
If the local node $v$ has not the priority on the too far node $w$
the lists in which $w$ appears are ignored (Line~\ref{l:order-dbm}).
At the opposite side of the group, node $w$ keeps the list containing $v$ but
the end of its ordered list of ancestor's sets will be truncated (meaning that
$v$ and $w$ will not belong to the same group).
Indeed, after the too far nodes have been all examined, the list of ancestors is
computed again (Lines~\ref{l:compute2b}-\ref{l:compute2e}) and is truncated
(Line~\ref{l:truncate}) in order to delete the too far nodes (these remaining
too fare nodes have less priority than $v$).

In order to not include a node in a view while it could be rejected later, a
\emph{quarantine mechanism} is used. The quarantine period of a node willing to
enter in a group is equal to \texttt{Dmax} timers. Each time a computation is
done (and then the new node progresses in the group), its quarantine period
decreases. Since the group diameter is less than or equal to \texttt{Dmax}, any
conflict would have been detected before the new node enters into a
view. Moreover, if a member of the group accepts the new node, then all the
members will accept it.

Finally, the priority is updated. When using oldness in the group, the priority
is increased if the node is not in a group. If the node is in a group, the
priority remains stable. The group priority is the smallest priority of its members.

The procedure \texttt{compute()} is given below (a complete implementation with
the detailed algorithm is available on-line, see reference in
Footnote~\pageref{refonline} page~\pageref{refonline}).

\begin{footnotesize}
\renewcommand{\AlgTextName}{{\AlgStyName Procedure}}
\begin{algorithm*}{compute() on node $v$}
\AlgSetWidthCom{0.45\linewidth}
  \C{Checking the received lists}
  \FORALL{ $\texttt{list}_u$ in $\texttt{msgSet}$ }
    \I delete marked nodes except $\underline{v}$ in $\texttt{list}_u$%
       \label{l:mn}%
 		   \CL{Marked nodes are only useful between neighbors.}
 	  \IF{ $\neg$ goodList($\texttt{list}_u$) }\label{l:good}
       \CL{List of $u$ cannot be used;}
   		 \I replace $\texttt{list}_u$ by $(\underline{u})$ in \texttt{msgSet} \label{l:badlist}\label{l:sm} \CL{this
  list is ignored but the sender is kept.}
		\ENDIF\label{l:size} \CL{Now, incoming lists cannot be larger than $\texttt{Dmax}$.}
		\IF{ $u \not\in \texttt{view}_v$ \AND $\neg$ compatibleList($\texttt{list}_u$) }  \label{l:cond}
	  \CL{$u$ is new, but its list cannot be accepted;}
   		\I replace $\texttt{list}_u$ by $(\underline{\underline{u}})$ in
  \texttt{msgSet} \CL{$u$ is denoted as an incompatible neighbor} 
    \label{l:cond-dbm} \label{l:dbm1}
		\ENDIF
	\ENDFOR
  \C{Computing the list of ancestors' sets of $v$.}
	\I \texttt{list}$_v$ \= $(v)$  \label{l:compute1b}
  \FORALL{ $\texttt{list}_u \in \texttt{msgSet}$ }
	  \I $\texttt{list}_v$ \= $\mathrm{ant}(\texttt{list}_v, \texttt{list}_u)$
       \CL{Computation using the $\mathrm{ant}$ $r$-operator.}
  \ENDFOR \label{l:compute1e}
  \C{Removal of incoming lists containing too far nodes (after $\mathrm{ant}$
  computation, \emph{$\texttt{list}_v$} cannot be larger than \emph{$\texttt{Dmax} +1$})}
  \IF{ $s(\texttt{list}_v) = \texttt{Dmax} + 2$ } 	  \CL{The list is too long.}
    \FORALL{ $w$ at position $\texttt{Dmax}+1$ in $\texttt{list}_v$ } \CL{Scanning too far nodes.}
      \IF{ $w$ has the priority compared to $v$} \label{l:order} 
			  \CL{Far node $w$ has the priority.}
        \FORALL{$\texttt{list}_u \in \texttt{msgSet}$} \CL{Looking for lists that provided $w$;}
          \IF{ $w$ is at position $\texttt{Dmax}$ } \CL{they contain $w$ in their last place.}
 					  \I replace $\texttt{list}_u$ by $(\underline{\underline{u}})$ in
            \texttt{msgSet} \label{l:dbm2}
						\label{l:order-dbm}\CL{The neighbor that provided $w$ is ignored.}
				  \ENDIF
        \ENDFOR
      \ENDIF
    \ENDFOR
		\C{Computing $\texttt{list}_v$ again, without the incoming lists that contained too far
    nodes with priority.} 
    \I $\texttt{list}_v$ \= $(v)$ \label{l:compute2b}
    \FORALL{ $\texttt{list}_u$ in \texttt{msgSet} }
	    \I $\texttt{list}_v$ \= $\mathrm{ant}(\texttt{list}_v, \texttt{list}_u)$
    \ENDFOR \label{l:compute2e}
	  \I keeping up to $\texttt{Dmax}+1$ first elements in $\texttt{list}_v$
    \label{l:truncate}
    \CL{Deleted too far nodes have not the priority.}
  \ENDIF
  \I Update quarantines: quarantine of new nodes is \texttt{Dmax}, non null quarantine of
  others decreases by 1
	\I $\texttt{view}_v$ \= non marked nodes in $\texttt{list}_v$ with null quarantine \label{l:view}
	\I Update priorities: priority of nodes increase only when they are not in a group
%
%
\end{algorithm*}
\renewcommand{\AlgTextName}{{\AlgStyName Function}}
\begin{algorithm*}{goodList(\texttt{list})}
\IF{ $v$ or $\underline{v}$ are in $\texttt{list}.1$ \AND
	  $s(\texttt{list}) \leq \texttt{Dmax} +1$ \AND $\emptyset \notin
	  \texttt{list}$ }
  \RETURN true
\ELSE
\RETURN false
\end{algorithm*}
\begin{algorithm*}{compatibleList(\texttt{list})}
\I\AlgTextIF $s(\texttt{list}_v) + s(\texttt{list}) \leq \texttt{Dmax}+1$ \OR \\
	\noindent\mbox{}\hspace{\AlgInd}
  $\exists i \in \{0,\ldots,s(\texttt{list}_v)\}$,
  $\texttt{list}_v.i \subseteq \texttt{list}.1 \wedge \min\left(s(\texttt{list}_v) + s(\texttt{list}) +1-i,
  s(\texttt{list}) + 1 + i/2\right)  \leq \texttt{Dmax}$
	    \RETURN true \CL{Refer to Proposition~\ref{p:compatible}.}
\ELSE
\RETURN false
\end{algorithm*}

\end{footnotesize}

\section{Proofs}\label{s:proof}

We first focus on the self-stabilizing property of our algorithm. We show that
assuming a fixed topology, the system converges in finite time to an execution
satisfying the statements in Section~\ref{s:spec}, \ie $\Pi_S \wedge \Pi_A
\wedge \Pi_M$ is an attractor.  Next, we prove that, assuming topological
changes preserving the maximal distance condition over the groups, then
continuity is preserved, \ie $\Pi_T \Rightarrow \Pi_C$.

\subsection{Stabilization}

In this section, we prove that our protocol is self-stabilizing by showing that
$\Pi_S$ and $\Pi_A$ and $\Pi_M$ are attractors---%
Propo\-sitions~\ref{p:safe}, \ref{p:agree} and \ref{p:max}, respectively.

We begin by showing that eventually lists will become correct
(Propositions~\ref{p:dmax} and \ref{p:exist}).
We first prove that any execution cannot remain infinitely with configurations
having lists larger than \texttt{Dmax}. We denote by $e_\texttt{Dmax}$ the
suffix of an execution $e$ such that, for any configuration $c \in
e_\texttt{Dmax}$, for any node $v \in V$, the size of \texttt{list}$_v$ is
smaller than or equal to $\texttt{Dmax}+1$.

\begin{proposition}[Dmax]\label{p:dmax}
On a fixed topology, any execution $e$ reaches in finite time a suffix
$e_\texttt{Dmax}$.
\end{proposition}

\begin{proof}
Starting from configuration $c_1$, the system will reach in finite time a
configuration in which every node has computed its list after expiration of its
timer. After such a computation, the size of the lists is bounded by
$\texttt{Dmax}+1$ (because it is truncated at the $\texttt{Dmax}+1$ position,
Line~\ref{l:truncate}).
\end{proof}

Starting from this proposition, we now prove that any execution cannot remain
infinitely with configurations having a non existing node in a list. We denote
by $e_\texttt{exist}$ the suffix of an execution $e$ such that, for any
configuration $c \in e_\texttt{exist}$, for any node $v\in V$, every node $u \in
\texttt{list}_v^c$ satisfies $u \in V$.

\begin{proposition}[Exist]\label{p:exist}
On a fixed topology, any execution $e$ reaches in finite time a suffix
$e_\texttt{exist}$.
\end{proposition}

\begin{proof}
Let $c \in e_\texttt{Dmax}$ be a configuration (Proposition~\ref{p:dmax}).
Let $u$ be a node label such that $u \not\in V$ and denote by $U_k^c$ the set of
nodes having $u$ in their list at position $k$ in configuration $c$.  Consider
the function $\phi(c)$ defined by $\phi(c) = \min \{ k\in\mathbb{N}, U_k^c \neq
\emptyset \}$ and $\phi(c) = \infty$ if $\forall k \in \mathbb{N}, U_k^c =
\emptyset$. We prove that $\phi$ is continuously growing along the execution to
be eventually equal to infinity forever.

Consider a node $v$ in $U^{c}_{\phi(c)}$: $v$ contains $u$ at position
$\phi(c)$ in its computed list and no node in configuration $c$ contains $u$
at a smaller position in its computed list. Until the next expiration of its
timer, $v$ cannot receive a list containing $u$ in a smaller position than
$\phi(c)$.
Hence, the system will reach in finite time a configuration in which the node
$v$ has computed a new list that does not contain $u$ at a position smaller than
$\phi(c)+1$. After a timer (fair channel Hypothesis), the system reaches in
finite time a configuration in which the neighbors of $v$ have received this
list.

After finite time, any node $v \in U^{c}_{\phi(c)}$ will do the same. The system
then reaches in finite time after configuration $c$ a configuration $c'$ in
which $U^{c'}_{\phi(c)}$ is empty, meaning that $\phi(c) < \phi(c')$.

By iteration, $\phi$ is growing along the execution.  Since the size of the
lists is bounded by $\texttt{Dmax}+1$ (Proposition~\ref{p:dmax}), there exists a
configuration $c''$ reached in finite time after $c$ in which $\phi(c'') =
\infty$, meaning that $u$ does not appear anymore in the computed lists of the
nodes forever.
\end{proof}


Next, we establish the connection between marked nodes in the algorithms and
subgraphs (Propositions~\ref{p:nopropagation}, \ref{p:propagation}, \ref{p:dme}
and \ref{p:subgraphs}).
We call \emph{double-marked edge} an edge $(u,v)$ such that either $u$
double-marks $v$ or $v$ double-marks $u$ (denoted by $\underline{\underline{u}}$
in the algorithm).  The following proposition is a consequence of the
double-marked edge technique. A node $v$ double-marks its neighbor $u$ only if
the list sent by $u$ cannot be accepted by $v$ (Lines~\ref{l:dbm1} and
\ref{l:dbm2}). In this case, node $v$ will ignore the list sent by
$u$. Reciprocally, if $u$ has been double-marked by $v$, $u$ will detect an
asymmetric link ($u$ does not appear in the list it received after
Line~\ref{l:mn}) and only the identity of $v$ will be kept by $u$, the rest of
the list of $v$ will be ignored (Line~\ref{l:sm}).

\begin{proposition}[No Propagation]\label{p:nopropagation}
  Let $u$ and $v$ be two vertices of $G$ and suppose that, in any execution $e$,
  there exists a configuration $c_e$ from which any path from $u$ to $v$ in $G$
  contains a double-marked edge. Then $u$ will eventually disappear from
  $\texttt{list}_v^{c}$ and $v$ will eventually disappear from $\texttt{list}_u^{c}$.
\end{proposition}

The following proposition is a consequence of the \emph{ant} computation (see
Section~\ref{s:ant}). It propagates nodes identities (providing there is no
edge-marking technique for limiting it) \cite{DT03,JACIC06}.

\begin{proposition}[Propagation]\label{p:propagation}
  Let $u$ and $v$ be two vertices of $G$ and suppose that, in any execution $e$,
  there exists a configuration $c_e$ from which there exists a path from $u$ to $v$
  in $G$ without double-marked edge. Then $\texttt{list}_v^{c}$ will eventually
  contain $u$ and $\texttt{list}_u^{c}$ will eventually contain $v$.
\end{proposition}

\begin{proposition}[Double-marked edge]\label{p:dme}
  Suppose that $d(u,v) > \texttt{Dmax}$. Then any execution admits a suffix
  $e_\texttt{edge}$ such that, for any configuration $c \in e_\texttt{edge}$,
  there is a double-marked edge on any path from $u$ to $v$.
\end{proposition}

\begin{proof}
  Let $v$ and $w$ two nodes of $G$ such that $d(v,w) = \texttt{Dmax} + 1$. Without loss
  of generality, we suppose that $pr(w) < pr(v)$. Suppose that there exists a
  path from $v$ to $w$ that does not contain any double-marked edge. By
  Proposition~\ref{p:propagation}, there exists a neighbor $u$ of $v$ such that
  $u$ sends to $v$ a list containing $w$. The size of this list is larger than
  \texttt{Dmax}. There are two cases:
  \newline \noindent 
  ($i$) $u \not\in \texttt{view}_v$. In this case, $\texttt{list}_u$ is
    replaced by $(\underline{\underline{u}})$.
  \newline \noindent 
  ($ii$) $u \in \texttt{view}_v$. In this case, $v$ computes a list using the
  one sent by $u$. Since $d(u,v) > \texttt{Dmax}$, the resulting list is too
  long. Since $pr(w) < pr(v)$, the computation will be done again without the
  list provided by $u$, which will be replaced by
  $(\underline{\underline{u}})$.

  In the two cases, $u$ is double-marked by $v$. Hence, any path from $u$ to $v$
  will eventually contains a double-marked edge.
\end{proof}

Let denote by $H_v^c(V_{H_v},E_{H_v})$ the subgraph of $G(V,E)$ defined in the
configuration $c$ by: for any node $u$ in $V_{H_v}$, $v \in \texttt{list}_u^c$.
Such a subgraph is composed of vertices containing $v$ in their list. We prove
that eventually $H_u$ and $H_v$ are distinct when $d(u,v) > \texttt{Dmax}$.

\begin{proposition}[Subgraphs]\label{p:subgraphs}
  Suppose that $d(u,v) > \texttt{Dmax}$. Then any execution admits a suffix
  $e_\texttt{subgraph}$ such that, for any configuration $c \in e_\texttt{subgraph}$,
  $H_u$ and $H_v$ are distinct subgraphs.
\end{proposition}

\begin{proof}
  By Proposition~\ref{p:dme}, there exists a suffix $s_1$ such that any path
  from $u$ to $v$ contains a double-marked edge. By
  Proposition~\ref{p:nopropagation}, there exists a suffix $s_2$ included in
  $s_1$ such that for any configuration $c$ in this suffix, $u \not\in
  \texttt{list}_v^c$ and $v \not\in\texttt{list}_u^c$. Then $u \not\in H_v$ and
  $v \not\in H_u$.

  Let consider a node $w$ such that $w \in H_v$ and $w\in H_u$. Then there
  exists at least one path from $u$ to $v$ containing $w$. The length of such a
  path is larger than \texttt{Dmax}. Then, by Proposition~\ref{p:dme}, it
  admits a double-marked edge, either on the subpath from $u$ to $w$ or from the
  subpath from $w$ to $v$.

  Now, let consider all the paths from $u$ to $v$ containing $w$; they all
  contain a double-marked edge. Suppose that for one path $P_1$, this double-marked edge is between $w$ and $v$ and for a second path $P_2$, it is between
  $u$ and $w$. Then, by considering edges of $P_1$ from $u$ to $w$ and edges of
  $P_2$ from $w$ to $v$, we obtain a path from $u$ to $v$ without any double-marked edge, which is a contradiction. Then, all paths from $u$ to $v$
  containing $w$ admit a double-marked edge, and this edge is always between $u$
  and $w$ or always between $w$ and $v$. Thus, $w$ cannot belong to both $H_u$
  and $H_v$, meaning that there is no node $w$ such that $w \in H_u$ and $w\in
  H_v$.

  Hence, any execution reaches a suffix such that, for any configuration $c$ in
  this suffix, $H_u^c$ and $H_v^c$ are distinct.
\end{proof} 


The preceding propositions give the Agreement.  Consider any execution
$e_\texttt{subgraphs}$.  Denote by $e_\texttt{agree}$ the suffix of an execution
$e$ such that $\Pi_A(c)$ holds for any configuration $c\in e_\texttt{agree}$,
that is $V_{H_v} = \texttt{view}_w^c$ for any $w \in H_v$.
The following proposition is given by Propositions~\ref{p:subgraphs},
\ref{p:propagation} and \ref{p:nopropagation}.

\begin{proposition}[Agreement]\label{p:agree}
  On a fixed topology, any execution $e$ reaches in finite time a suffix
  $e_\texttt{agree}$.
\end{proposition}

\begin{proof}
  By Proposition~\ref{p:subgraphs}, for any execution, there exists a suffix
  such that, for any nodes $u$ and $v$ in $G$, if $d(u,v) > \texttt{Dmax}$,
  then the subgraphs $H_u$ and $H_v$ are distinct. Consider now two nodes $w$
  and $v$ such that $w$ belongs to $H_v$

  By Proposition~\ref{p:propagation}, for any execution, there exists a suffix
  such that, for any configuration $c$ in this suffix, the identities of $H_v$
  will be in $\texttt{list}_w^c$.

  By Proposition~\ref{p:nopropagation}, for any execution, there exists a suffix
  such that, for any configuration $c$ in this suffix, the $\texttt{list}_w^c$
  contains only vertices of $H_v$.

  After the end of the quarantine period, all the nodes in $\texttt{list}_w$
  belong to $\texttt{view}_w$.  Then the system reaches a suffix in which all
  the nodes of $H_v$ and only these nodes appear in $\texttt{view}_w$, for any
  vertex $w \in H_v$. Hence, $\texttt{view}_v^c = \texttt{view}_w^c =
  \Omega_v^c$. This gives $\Pi_A$.
\end{proof}

Now we have the agreement, there is a connection between subgraphs and
groups. We then prove the Safety.
Consider any execution $e_\texttt{agree}$. 
Denote by $e_\texttt{safe}$ the suffix of an execution $e$ such that $\Pi_S(c)$
holds for any configuration $c\in e_\texttt{safe}$. The following proposition is
a consequence of Prop.~\ref{p:subgraphs}.

\begin{proposition}[Safety]\label{p:safe}
On a fixed topology, any execution $e$ reaches in finite time a suffix $e_\texttt{safe}$.
\end{proposition}

\begin{proof}
  By Proposition~\ref{p:subgraphs}, for any execution and any nodes $u$ and $v$
  in $G$ satisfying $d(u,v) > \texttt{Dmax}$, the subgraphs $H_u$ and $H_v$ will
  eventually be distinct. Hence, for any execution, there exists a suffix
  $e_\texttt{safe}$ such that, for any configuration $c \in e_\texttt{safe}$,
  for any vertex $v$ in $G$, $\text{Diam}(H_v^c) \leq \texttt{Dmax}$.

  Then, by Proposition~\ref{p:agree}, we have $\max_{x,y \in \Omega_v^c}
  d_{\Omega_v^c}(x,y) \leq \texttt{Dmax}$. This gives $\Pi_S$.
\end{proof}


We consider any execution $e_\texttt{agree}$.  In order to prove the maximality
property, we introduce the following definitions.  An edge $(u,v)$ is
\emph{internal} in a given configuration $c$ if $\Omega_u^c = \Omega_v^c$.  In
the converse case ($\Omega_u^c \neq \Omega_v^c$), it is \emph{external}.
An external edge involves double-marked nodes and it is then not propagated by
the algorithm (marked nodes are deleted, see line~\ref{l:mn} in Procedure
\texttt{compute()}).
We denote by $nee$ (resp. $ndg$) the function defined on $\mathcal{C}$ that
returns the \underline{n}umber of \underline{e}xternal \underline{e}dges in a
given configuration (resp. the \underline{n}umber of \underline{d}istinct
\underline{g}roups in configuration $c$: $ndg(c) = |\{ \Omega_v^c, v \in V\}|$.

\begin{proposition}[Nee]\label{p:nee}
  If $nee$ is decreasing along a suffix $e_s$ of an execution $e$, $ndg$ is also
  decreasing along $e_s$.
\end{proposition}

\begin{proof}
  Let $(u,v)$ be an external edge in a configuration $c_i$ and assume that it is
  an internal edge in configuration $c_{i+1}$. This means that $\Omega_u^{c_i}
  \neq \Omega_v^{c_i}$ and $\Omega_u^{c_{i+1}} = \Omega_v^{c_{i+1}}$. Hence
  $nee(c_i) > nee(c_{i+1}) \Rightarrow ndg(c_i) > ndg(c_{i+1})$.
\end{proof}

We prove that any execution reaches in finite time a suffix in which the
function $nee$ does not increase. We denote by $e_\texttt{notincr}$ such a
suffix: $\forall c_i, c_{i+1} \in e_\texttt{notincr}$, $nee(c_{i+1}) \leq
nee(c_i)$.

\begin{proposition}[Not incr.]\label{p:notincr}
  On a fixed topology, any execution $e$ reaches in finite time a suffix
  \emph{$e_\texttt{notincr}$}.
\end{proposition}

\begin{proof}
  Let $c \in e_\texttt{agree}$ be a configuration (Proposition~\ref{p:agree}).
  Let $(u,v)$ be an internal edge in configuration $c$. Then we have $\Omega_u^c
  = \Omega_v^c$ and $u$ is in $\texttt{list}_v^c$.
  In order $(u,v)$ becomes an external edge, one of its extremity (say $v$)
  would have double-marked the other (in Procedure \texttt{compute()}). But this
  cannot happen after the \texttt{goodList} test (line~\ref{l:good}) because $c
  \in e_\texttt{subgraphs}$. This cannot happen after the \texttt{compatibleList}
  test (line~\ref{l:cond}) because $u$ is in already in \texttt{view}$_v^c$.
\end{proof}

Now, we prove that any execution reaches in finite time a suffix in which the
function $nee$ is decreasing while $\Pi_M$ is not true. We denote by
$e_\texttt{decr}$ such a suffix: $\forall c_i \in e_\texttt{decr}$, $\Pi_M(c_i)
\vee \exists c_j \in e_\texttt{decr}$, $i < j$ and $nee(c_i) > nee(c_j)$.

 \begin{proposition}[Decreasing]\label{p:decr}
 On a fixed topology, any execution $e$ reaches in finite time a suffix $e_\texttt{decr}$.
 \end{proposition}

\begin{proof}
  Let $c \in e_\texttt{notincr}$ be a configuration
  (Proposition~\ref{p:notincr}). Starting from such a configuration, the $nee$
  function cannot increase.
  Suppose that $\Pi_M$ is not true in $c$. Then, by definition of $\Pi_M$, there
  exists two neighbors nodes $x$ and $y$ with different views that could merge
  their groups without breaking $\Pi_S$. By fair channel hypothesis, a timer
  later the system reaches a configuration $c'$ in which $x$ (resp. $y$) has
  received the list sent by $y$ (resp. $x$).

  Without loss of generality, suppose that $\Omega_x$ has the smallest priority
  among all the subgraphs that can merge, and $\Omega_y$ has the smallest priority
  among all the groups that can merge with $\Omega_x$.

  During the \texttt{compute()} Procedure on $x$ and $y$, the \texttt{goodList}
  tests are true because $c' \in e_\texttt{notincr}$ and then $c' \in
  e_\texttt{safe}$. The \texttt{compatibleList} test is true on both $x$ and $y$
  because they cannot have change their list since configuration $c$. Hence we
  obtain: $x \in \texttt{list}_y$ and $y \in \texttt{list}_x$.

  Since $\Omega_y$ has the smallest priority among the neighbors of $\Omega_x$, no member
  of $\Omega_x$ can receive a message from a group with a smallest
  priority. Therefore $x$ will never receive and then will never send to $y$ a
  list with a too far node with a smallest priority than $y$ one's. Hence $y$
  will never double-mark $x$ and $x$ will remain in the list of $y$.

  Similarly, since $\Omega_x$ has the smallest priority among the groups that can
  merge, no member of $\Omega_y$ can receive a message from a group with a smallest
  priority. Therefore $y$ will never receive and then will never send to $x$ a
  list with a too far node with a smallest priority than $x$ one's. Hence $x$
  will never double-mark $y$ and $y$ will remain in the list of $x$.

  After \texttt{Dmax} timer, the list of $y$ (resp. $x$) has reached any $u \in
  \Omega_x$ (resp. $\Omega_y$) thanks to the fair channel Hypothesis. Moreover the
  quarantine of these new members reaches $0$ and they are now included in
  $\texttt{view}_u$. Thus, the edge $(x,y)$ becomes an internal edge.

  Hence, starting from configuration $c$ with $\neg\Pi_M(c)$, the system reaches
  in finite time a configuration $c''$ with $nee(c) > nee(c'')$.
\end{proof}

The following proposition is given by Propositions~\ref{p:nee}, \ref{p:notincr}
and ~\ref{p:decr}; it shows that any execution reaches in finite time a suffix
in which $\Pi_M$ is true. We denote by $e_\texttt{max}$ such a suffix.

\begin{proposition}[Maximality]\label{p:max}
  On a fixed topology, any execution $e$ reaches in finite time a suffix
  $e_\texttt{max}$.
\end{proposition}

\begin{proof}
  By Prop.~\ref{p:notincr}, the execution reaches a suffix $e_\texttt{notincr}$
  such that the $nee$ function will no more increase.  By Prop.~\ref{p:decr},
  the execution reaches a suffix $e_\texttt{decr}$ such that the $nee$ function
  decreases while $\Pi_M$ is not true.  Hence, while $\Pi_M$ is false, the
  number of external edges will eventually decrease. By Prop.~\ref{p:nee}, this
  means that the number of subgraphs will eventually decrease while $\Pi_M$ is
  false. Since the graph is finite, the number of subgraphs cannot decrease
  infinitely and $\Pi_M$ will eventually become true.
\end{proof}

\subsection{Best-effort requirement}

We now consider the dynamic of the network.  We show that if the continuity
property is violated into a group, then there exists a pair of nodes belonging
to that group such that the distance between them is larger that
\texttt{Dmax}. The following technical proposition justifies the compatibleList
test.

\begin{proposition}[Compatible lists]\label{p:compatible}
Let $v$ be a node having the list $(a_v^0,a_v^1,$ $\ldots,$ $a_v^p),\ (p\geq 0)$ and assume
that its neighbor $w$ sends the list $(a_w^0,a_w^1, \ldots, a_w^q),\ (q\geq 0)$.
Then, the diameter of the group of $v$ after $v$ accepts $w$ remains smaller than or equal to
\texttt{Dmax} if and only if there exists $i \in
\{0,\ldots,p\}$ such that $w$ is neighbor of all the nodes belonging to $a_v^i$
and either $p-i+1+q \leq \texttt{Dmax}$ or $i/2+q+1 \leq \texttt{Dmax}$.
\end{proposition}

\begin{proof}
  Let $c \in e_\texttt{safe}$ be a configuration (Proposition~\ref{p:safe}).
  Let $w$ be the first node of $\Omega_w^c$ for which the list of ancestor's sets is
  received by $v$. Then, the only external edges between $\Omega_v^c$ and $\Omega_w^c$
  known by $v$ are those joining $w$ (external edges are not propagated). Hence,
  without loss of generality, assume that only these external edges exist
  between the groups.

  \noindent ($\Rightarrow$) Assume that the conditions are fulfilled.
  Let $u \in a_v^k$ and $u' \in a_w^l$ be two nodes in the lists of $v$ and $w$
  respectively. There exists at most two families of shortest paths from $u$ to
  $u'$, depending on the external edge used to reach $w$.
  Let $P_1$ be a path that includes the edge $(v,w)$. It starts from $u$ and
  joins $v$ by $k$ edges in the group of $v$, joins $w$ by the edge $(u,v)$ and
  then reaches $u'$ by $l$ edges in the group of $u$.
  Let $P_2$ be a path from the second family. It starts from $u$ and joins a
  node $v' \in a_v^i$ by $|k-i|$ internal edges in the group of $v$, then joins
  $w$ by the edge $(v',w)$ and then reaches $u'$ by $l$ internal edges in the
  group of $u$.

  The length of $P_1$ is bounded by $k + 1 + q$.  But since $P_1$ is a shortest
  path, it is shorter to reach $u'$ from $u$ by joining a node of $a_v^0$ (\ie
  $v$) than by joining a node of $a_v^i$ (such as $v'$). Hence we have $k \leq
  i/2$ and the length of $P_1$ is bounded $i/2 + 1 + q$, which is smaller than
  $\texttt{Dmax}$ by hypothesis.
  The length of $P_2$ is bounded by $p - i + 1 + q$, which is also smaller than
  $\texttt{Dmax}$ by assumption.

  Hence, for any node $u$ and $u'$ belonging to the group of $v$ and $w$
  respectively, there exists a path from $u$ to $u'$ with less than
  $\texttt{Dmax}$ edges. The list of $w$ is then compatible with the list of
  $v$, and can then be accepted by $v$.

  \noindent ($\Leftarrow$) Assume by contradiction that the conditions are not
  fulfilled and that $v$ accepts the list of $w$, \ie $v$ includes the list of
  $w$ by computing its new list with \texttt{ant}---refer to Lines~$14-16$ of
  Procedure \texttt{compute()}.  That means that the list of $w$ is
  compatible---refer to Lines~$6-8$---, which contradicts the assumption.  Then
  the nodes of $\texttt{list}_w^c$ will be propagated in the lists of nodes of
  $\texttt{list}_v^c$ and reciprocally. But at least one node $u \in
  \texttt{list}_v^c$ will see that a node $u' \in \texttt{list}_w^c$ is too
  far from it and reciprocally. Either $u$ or $u'$ will reject the lists of its
  neighbors that contain the too far node (depending on the priority between $u$
  and $u'$) and either the group of $v$ or the group of $w$ splits (when a
  neighbor is rejected by $u$, it disappears from $\texttt{list}_u$, and then
  from $\texttt{view}_u$; it is then no more in $H_v$).
\end{proof}

\begin{proposition}
  For any execution $e$, for any configuration $c_i$ in $e$, $\Pi_T(c_i,c_{i+1})
  \Rightarrow \Pi_C(c_i, c_{i+1})$.
\end{proposition}

\begin{proof}
  Suppose that there exists a configuration $c_i$ and a node $v$ such that
  $\texttt{view}_v^{c_i} \not\subseteq \texttt{view}_v^{c_{i+1}}$. Then there
  exists a node $u$ such that $u \in \texttt{view}_v^{c_i}$ and $u \not \in
  \texttt{view}_v^{c_{i+1}}$.
  This cannot happen after $u$ or $v$ has added a new node in its view, thanks
  to the quarantine mechanism. This can only happen because either $u$ or $v$
  removed a node from their views.

  Without loss of generality, suppose that $v$ removed a node $x$: $x \in
  \texttt{view}_v^{c_i}$ and $x \not\in \texttt{view}_v^{c_{i+1}}$.
  If $x \not\in \texttt{view}_v^{c_{i+1}}$, then ($i$) the quarantine of $x$ is
  not null or ($ii$) $x$ is not in $\texttt{list}_v^{c_{i+1}}$ or ($iii$) $x$ is
  marked in $\texttt{list}_v^{c_{i+1}}$ (Line~\ref{l:view} in Procedure
  \texttt{compute()}).

\noindent 
  ($i$) The first case is exclude because $x$ was already in
  $\texttt{view}_v^{c_i}$.

\noindent 
  ($ii$) In the second case, if $v$ has not received the message of $x$ while it
  received it before, then $x$ left the neighborhood of $v$.
  Then, in configuration $c_{i+1}$, there is not path from $x$ to
  $v$ with only nodes of $\Omega_v^{c_i}$ and $d_{\Omega_v^{c_i}}^{c_{i+1}}(x,v)
  = +\infty$. Thus $\neg\Pi_T(c_i, c_{i+1})$ (\emph{a neighbor left}).

\noindent 
  ($iii$) In the third case, if $x$ is simple marked, its list is not good while
  it was in configuration $c_i$, which is exclude (Line~\ref{l:good}). If $x$ is
  double-marked, this cannot happen after the compatibleList test
  (Line~\ref{l:cond-dbm}) because $x$ was in $\texttt{view}_v^{c_i}$. If this
  happened after Line~\ref{l:order-dbm}, then $x$ sent a list with a too far
  node $y$ having priority on $v$. If $y \not \in \Omega_v^{c_i}$, then $y \not
  \in \texttt{view}_v^{c_i}$. Then the quarantine of $y$ is not null and no node
  of $\Omega_v^{c_i}$ has admitted $y$ in its view. Therefore, thanks to
  Prop.~\ref{p:compatible}, $y$ would have never been propagated inside
  $\Omega_v^{c_i}$ until $v$, because of the compatibleList test
  (Line~\ref{l:cond}). Finally, if $y \in \Omega_v^{c_i}$, then the distance
  from $y$ to $v$ in configuration $c_{i+1}$ is larger than \texttt{Dmax}:
  $d_{\Omega_v^{c_i}}^{c_{i+1}}(x,v) > \texttt{Dmax}$ and $\neg \Pi_T(c_i,
  c_{i+1})$. 
\end{proof}

\section{Conclusion}
\label{s:conclu}

This paper introduces the best effort requirement to complete the
self-stabilization for designing algorithm in dynamic networks.
To illustrate this approach, a new problem inspired from VANET has been
specified: the \emph{Dynamic Group Service}.  A best effort distributed protocol
called \texttt{GRP} has been designed and proved for solving this problem in
message passing. The algorithm is self-stabilizing and fulfills a continuity
property whenever the dynamic allows it.
The protocol has been implemented and its performances studied by simulation
(see reference in Footnote~\ref{refonline} page~\pageref{refonline}). We believe
that the best effort requirement is promising for building useful services in
dynamic networks.

\label{sec:biblio}

\newpage
\mbox{}
\vspace{1cm}
\appendix

\section{Omitted proofs}

\subsection{Proof of Proposition~\ref{p:dmax} (Dmax)}
\begin{proof}
 Starting from configuration $c_1$, the system will reach in finite time a
 configuration in which every node has computed its list after expiration of its
 timer. After such a computation, the size of the lists is bounded by
 $\texttt{Dmax}+1$ (because it is truncated at the $\texttt{Dmax}+1$ position,
 Line~\ref{l:truncate}).
 \end{proof}

\subsection{Proof of  Proposition~\ref{p:exist} (Exist)}

\begin{proof}
Let $c \in e_\texttt{Dmax}$ be a configuration (Proposition~\ref{p:dmax}).
Let $u$ be a node label such that $u \not\in V$ and denote by $U_k^c$ the set of
nodes having $u$ in their list at position $k$ in configuration $c$.  Consider
the function $\phi(c)$ defined by $\phi(c) = \min \{ k\in\mathbb{N}, U_k^c \neq
\emptyset \}$ and $\phi(c) = \infty$ if $\forall k \in \mathbb{N}, U_k^c =
\emptyset$. We prove that $\phi$ is continuously growing along the execution to
be eventually equal to infinity forever.

Consider a node $v$ in $U^{c}_{\phi(c)}$: $v$ contains $u$ at position
$\phi(c)$ in its computed list and no node in configuration $c$ contains $u$
at a smaller position in its computed list. Until the next expiration of its
timer, $v$ cannot receive a list containing $u$ in a smaller position than
$\phi(c)$.
Hence, the system will reach in finite time a configuration in which the node
$v$ has computed a new list that does not contain $u$ at a position smaller than
$\phi(c)+1$. After a timer (fair channel Hypothesis), the system reaches in
finite time a configuration in which the neighbors of $v$ have received this
list.

After finite time, any node $v \in U^{c}_{\phi(c)}$ will do the same. The system
then reaches in finite time after configuration $c$ a configuration $c'$ in
which $U^{c'}_{\phi(c)}$ is empty, meaning that $\phi(c) < \phi(c')$.

By iteration, $\phi$ is growing along the execution.  Since the size of the
lists is bounded by $\texttt{Dmax}+1$ (Proposition~\ref{p:dmax}), there exists a
configuration $c''$ reached in finite time after $c$ in which $\phi(c'') =
\infty$, meaning that $u$ does not appear anymore in the computed lists of the
nodes forever.
\end{proof}

\subsection{Proof of Proposition~\ref{p:dme} (Double-marked edge)}

\begin{proof}
  Let $v$ and $w$ two nodes of $G$ such that $d(v,w) = \texttt{Dmax} + 1$. Without loss
  of generality, we suppose that $pr(w) < pr(v)$. Suppose that there exists a
  path from $v$ to $w$ that does not contain any double-marked edge. By
  Proposition~\ref{p:propagation}, there exists a neighbor $u$ of $v$ such that
  $u$ sends to $v$ a list containing $w$. The size of this list is larger than
  \texttt{Dmax}. There is two cases.
  (i) $u \not\in \texttt{view}_v$. In this case, $\texttt{list}_u$ is
    replaced by $(\underline{\underline{u}})$.
    (ii) $u \in \texttt{view}_v$. In this case, $v$ computes a list using the
    one sent by $u$. Since $d(u,v) > \texttt{Dmax}$, the resulting list is too
    long. Since $pr(w) < pr(v)$, the computation will be done again without the
    list provided by $u$, which will be replaced by
    $(\underline{\underline{u}})$.
  In the two cases, $u$ is double-marked by $v$. Hence, any path from $u$ to $v$
  will eventually contains a double-marked edge.
\end{proof}

\subsection{Proof of Proposition~\ref{p:subgraphs} (Subgraphs)}

\begin{proof}
  By Proposition~\ref{p:dme}, there exists a suffix $s_1$ such that any path
  from $u$ to $v$ contains a double-marked edge. By
  Proposition~\ref{p:nopropagation}, there exists a suffix $s_2$ included in
  $s_1$ such that for any configuration $c$ in this suffix, $u \not\in
  \texttt{list}_v^c$ and $v \not\in\texttt{list}_u^c$. Then $u \not\in H_v$ and
  $v \not\in H_u$.

  Let consider a node $w$ such that $w \in H_v$ and $w\in H_u$. Then there
  exists at least one path from $u$ to $v$ containing $w$. The length of such a
  path is larger than \texttt{Dmax}. Then, by Proposition~\ref{p:dme}, it
  admits a double-marked edge, either on the subpath from $u$ to $w$ or from the
  subpath from $w$ to $v$.

  Now, let consider all the paths from $u$ to $v$ containing $w$; they all
  contain a double-marked edge. Suppose that for one path $P_1$, this double-marked edge is between $w$ and $v$ and for a second path $P_2$, it is between
  $u$ and $w$. Then, by considering edges of $P_1$ from $u$ to $w$ and edges of
  $P_2$ from $w$ to $v$, we obtain a path from $u$ to $v$ without any double-marked edge, which is a contradiction. Then, all paths from $u$ to $v$
  containing $w$ admit a double-marked edge, and this edge is always between $u$
  and $w$ or always between $w$ and $v$. Thus, $w$ cannot belong to both $H_u$
  and $H_v$, meaning that there is no node $w$ such that $w \in H_u$ and $w\in
  H_v$.

  Hence, any execution reaches a suffix such that, for any configuration $c$ in
  this suffix, $H_u^c$ and $H_v^c$ are distinct.
\end{proof}

\subsection{Proof of Proposition~\ref{p:agree} (Agreement)}

\begin{proof}
  By Proposition~\ref{p:subgraphs}, for any execution, there exists a suffix
  such that, for any nodes $u$ and $v$ in $G$, if $d(u,v) > \texttt{Dmax}$,
  then the subgraphs $H_u$ and $H_v$ are distinct. Consider now two nodes $w$
  and $v$ such that $w$ belongs to $H_v$

  By Proposition~\ref{p:propagation}, for any execution, there exists a suffix
  such that, for any configuration $c$ in this suffix, the identities of $H_v$
  will be in $\texttt{list}_w^c$.

  By Proposition~\ref{p:nopropagation}, for any execution, there exists a suffix
  such that, for any configuration $c$ in this suffix, the $\texttt{list}_w^c$
  contains only vertices of $H_v$.

  After the end of the quarantine period, all the nodes in $\texttt{list}_w$
  belong to $\texttt{view}_w$.  Then the system reaches a suffix in which all
  the nodes of $H_v$ and only these nodes appear in $\texttt{view}_w$, for any
  vertex $w \in H_v$. Hence, $\texttt{view}_v^c = \texttt{view}_w^c =
  \Omega_v^c$. This gives $\Pi_A$.
\end{proof}

\subsection{Proof of Proposition~\ref{p:safe} (Safety)}

\begin{proof}
  By Proposition~\ref{p:subgraphs}, for any execution and any nodes $u$ and $v$
  in $G$ satisfying $d(u,v) > \texttt{Dmax}$, the subgraphs $H_u$ and $H_v$ will
  eventually be distinct. Hence, for any execution, there exists a suffix
  $e_\texttt{safe}$ such that, for any configuration $c \in e_\texttt{safe}$,
  for any vertex $v$ in $G$, $\text{Diam}(H_v^c) \leq \texttt{Dmax}$.

  Then, by Proposition~\ref{p:agree}, we have $\max_{x,y \in \Omega_v^c}
  d_{\Omega_v^c}(x,y) \leq \texttt{Dmax}$. This gives $\Pi_S$.
\end{proof}

\subsection{Proof of Proposition~\ref{p:nee} (Nee)}

 \begin{proof}
   Let $(u,v)$ be an external edge in a configuration $c_i$ and assume that it is
   an internal edge in configuration $c_{i+1}$. This means that $\Omega_u^{c_i}
   \neq \Omega_v^{c_i}$ and $\Omega_u^{c_{i+1}} = \Omega_v^{c_{i+1}}$. Hence
   $nee(c_i) > nee(c_{i+1}) \Rightarrow ndg(c_i) > ndg(c_{i+1})$.
 \end{proof}

\subsection{Proof of proposition~\ref{p:notincr} (Not incr.)}

\begin{proof}
  Let $c \in e_\texttt{agree}$ be a configuration (Proposition~\ref{p:agree}).
  Let $(u,v)$ be an internal edge in configuration $c$. Then we have $\Omega_u^c
  = \Omega_v^c$ and $u$ is in $\texttt{list}_v^c$.
  In order $(u,v)$ becomes an external edge, one of its extremity (say $v$)
  would have double-marked the other (in Procedure \texttt{compute()}). But this
  cannot happen after the \texttt{goodList} test (line~\ref{l:good}) because $c
  \in e_\texttt{subgraphs}$. This cannot happen after the \texttt{compatibleList}
  test (line~\ref{l:cond}) because $u$ is in already in \texttt{view}$_v^c$.
\end{proof}

\subsection{Proof of Proposition~\ref{p:decr} (Decreasing)}

\begin{proof}
  Let $c \in e_\texttt{notincr}$ be a configuration
  (Proposition~\ref{p:notincr}). Starting from such a configuration, the $nee$
  function cannot increase.
  Suppose that $\Pi_M$ is not true in $c$. Then, by definition of $\Pi_M$, there
  exists two neighbors nodes $x$ and $y$ with different views that could merge
  their groups without breaking $\Pi_S$. By fair channel hypothesis, a timer
  later the system reaches a configuration $c'$ in which $x$ (resp. $y$) has
  received the list sent by $y$ (resp. $x$).

  Without loss of generality, suppose that $\Omega_x$ has the smallest priority
  among all the subgraphs that can merge, and $\Omega_y$ has the smallest priority
  among all the groups that can merge with $\Omega_x$.

  During the \texttt{compute()} Procedure on $x$ and $y$, the \texttt{goodList}
  tests are true because $c' \in e_\texttt{notincr}$ and then $c' \in
  e_\texttt{safe}$. The \texttt{compatibleList} test is true on both $x$ and $y$
  because they cannot have change their list since configuration $c$. Hence we
  obtain: $x \in \texttt{list}_y$ and $y \in \texttt{list}_x$.

  Since $\Omega_y$ has the smallest priority among the neighbors of $\Omega_x$, no member
  of $\Omega_x$ can receive a message from a group with a smallest
  priority. Therefore $x$ will never receive and then will never send to $y$ a
  list with a too far node with a smallest priority than $y$ one's. Hence $y$
  will never double-mark $x$ and $x$ will remain in the list of $y$.

  Similarly, since $\Omega_x$ has the smallest priority among the groups that can
  merge, no member of $\Omega_y$ can receive a message from a group with a smallest
  priority. Therefore $y$ will never receive and then will never send to $x$ a
  list with a too far node with a smallest priority than $x$ one's. Hence $x$
  will never double-mark $y$ and $y$ will remain in the list of $x$.

  After \texttt{Dmax} timer, the list of $y$ (resp. $x$) has reached any $u \in
  \Omega_x$ (resp. $\Omega_y$) thanks to the fair channel Hypothesis. Moreover the
  quarantine of these new members reaches $0$ and they are now included in
  $\texttt{view}_u$. Thus, the edge $(x,y)$ becomes an internal edge.

  Hence, starting from configuration $c$ with $\neg\Pi_M(c)$, the system reaches
  in finite time a configuration $c''$ with $nee(c) > nee(c'')$.
\end{proof}

\subsection{Proof of Proposition~\ref{p:compatible} (Compatible lists)}
\begin{proof}
  Let $c \in e_\texttt{safe}$ be a configuration (Proposition~\ref{p:safe}).
  Let $w$ be the first node of $\Omega_w^c$ for which the list of ancestor's sets is
  received by $v$. Then, the only external edges between $\Omega_v^c$ and $\Omega_w^c$
  known by $v$ are those joining $w$ (external edges are not propagated). Hence,
  without loss of generality, assume that only these external edges exist
  between the groups.

  \noindent ($\Rightarrow$) Assume that the conditions are fulfilled.
  Let $u \in a_v^k$ and $u' \in a_w^l$ be two nodes in the lists of $v$ and $w$
  respectively. There exists at most two families of shortest paths from $u$ to
  $u'$, depending on the external edge used to reach $w$.
  Let $P_1$ be a path that includes the edge $(v,w)$. It starts from $u$ and
  joins $v$ by $k$ edges in the group of $v$, joins $w$ by the edge $(u,v)$ and
  then reaches $u'$ by $l$ edges in the group of $u$.
  Let $P_2$ be a path from the second family. It starts from $u$ and joins a
  node $v' \in a_v^i$ by $|k-i|$ internal edges in the group of $v$, then joins
  $w$ by the edge $(v',w)$ and then reaches $u'$ by $l$ internal edges in the
  group of $u$.

  The length of $P_1$ is bounded by $k + 1 + q$.  But since $P_1$ is a shortest
  path, it is shorter to reach $u'$ from $u$ by joining a node of $a_v^0$ (\ie
  $v$) than by joining a node of $a_v^i$ (such as $v'$). Hence we have $k \leq
  i/2$ and the length of $P_1$ is bounded $i/2 + 1 + q$, which is smaller than
  $\texttt{Dmax}$ by hypothesis.
  The length of $P_2$ is bounded by $p - i + 1 + q$, which is also smaller than
  $\texttt{Dmax}$ by assumption.

  Hence, for any node $u$ and $u'$ belonging to the group of $v$ and $w$
  respectively, there exists a path from $u$ to $u'$ with less than
  $\texttt{Dmax}$ edges. The list of $w$ is then compatible with the list of
  $v$, and can then be accepted by $v$.

  \noindent ($\Leftarrow$) Assume by contradiction that the conditions are not
  fulfilled and that $v$ accepts the list of $w$, \ie $v$ includes the list of
  $w$ by computing its new list with \texttt{ant}---refer to Lines~$14-16$ of
  Procedure \texttt{compute()}.  That means that the list of $w$ is
  compatible---refer to Lines~$6-8$---, which contradicts the assumption.  Then
  the nodes of $\texttt{list}_w^c$ will be propagated in the lists of nodes of
  $\texttt{list}_v^c$ and reciprocally. But at least one node $u \in
  \texttt{list}_v^c$ will see that a node $u' \in \texttt{list}_w^c$ is too
  far from it and reciprocally. Either $u$ or $u'$ will reject the lists of its
  neighbors that contain the too far node (depending on the priority between $u$
  and $u'$) and either the group of $v$ or the group of $w$ splits (when a
  neighbor is rejected by $u$, it disappears from $\texttt{list}_u$, and then
  from $\texttt{view}_u$; it is then no more in $H_v$).
\end{proof}


\begin{thebibliography}{10}

\bibitem{APHV00}
A.D. Amis, R.~Prakash, and D.H.T. Vuaong.
\newblock Max-min $d$-cluster formation in wireless ad hoc networks.
\newblock In {\em IEEE INFOCOM}, pages 32--41, 2000.

\bibitem{ACM93}
Kenneth~P. Birman.
\newblock The process group approach to reliable distributed computing.
\newblock {\em Commun. ACM}, 36(12):37--53, 1993.

\bibitem{BEH04}
J.~Blum, A.~Eskandarian, and L.~Hoffman.
\newblock Challenges of intervehicle ad hoc networks.
\newblock {\em IEEE Transaction on Intelligent Transportation Systems,},
  5:347--351, 2004.

\bibitem{BDHY07}
O.~Brukman, S.~Dolev, Y.~Haviv, and R.~Yagel.
\newblock Self-stabilization as a foundation for autonomic computing.
\newblock In {\em The Second International Conference on Availability,
  Reliability and Security ({ARES})}, pages 991--998, Vienna, April 2007.

\bibitem{CKV01}
G.V. Chockler, I.~Keidar, and R.~Vitenberg.
\newblock Group communication specifications: a comprehensive study.
\newblock {\em ACM Computing Surveys}, 4(33):1--43, 2001.

\bibitem{DLV08}
A.~K. Datta, L.~L. Larmore, and P.~Vemula.
\newblock A self-stabilizing {O(k)}-time k-clustering algorithm.
\newblock {\em Computer Journal}, 2009.

\bibitem{JACIC06}
S.~Dela{\"e}t, B.~Ducourthial, and S.~Tixeuil.
\newblock Self-stabilization with r-operators revisited.
\newblock In {\em Journal of Aerospace Computing, Information, and
  Communication}, 2006.

\bibitem{Demirbas06}
Murat Demirbas, Anish Arora, Vineet Mittal, and Vinodkrishnan Kulathumani.
\newblock A fault-local self-stabilizing clustering service for wireless ad hoc
  networks.
\newblock {\em IEEE Trans. Parallel Distrib. Syst.}, 17(9):912--922, 2006.

\bibitem{D00}
S.~Dolev.
\newblock {\em Self-Stabilization}.
\newblock The MIT Press, 2000.

\bibitem{PODC95}
S.~Dolev and T.~Herman.
\newblock Superstabilizing protocols for dynamic distributed systems.
\newblock In {\em Proceedings of the fourteenth annual ACM symposium on
  Principles of distributed computing (PODC)}, page 255, New York, NY, USA,
  1995. ACM.

\bibitem{TMC06}
S.~Dolev, E.~Schiller, and J.L Welch.
\newblock Random walk for self-stabilizing group communication in ad hoc
  networks.
\newblock {\em IEEE Transactions on Mobile Computing}, 5(7):893--905, 2006.

\bibitem{SSS07}
B.~Ducourthial.
\newblock r-semi-groups: A generic approach for designing stabilizing silent
  tasks.
\newblock In {\em 9$^{th}$ Stabilization, Safety, and Security of Distributed
  Systems (SSS'2007)}, pages 281--295, Paris, novembre 2007.

\bibitem{DT03}
B.~Ducourthial and S.~Tixeuil.
\newblock Self-stabilization with path algebra.
\newblock {\em Theor. Comput. Sci.}, 293(1):219--236, 2003.

\bibitem{GS97}
R.~Guerraoui and A.~Schiper.
\newblock Software-based replication for fault-tolerance.
\newblock {\em IEEE Transaction on Computers}, 30(4):68--74, 1997.

\bibitem{DBLP:conf/icdcit/JhumkaK07}
Arshad Jhumka and Sandeep~S. Kulkarni.
\newblock On the design of mobility-tolerant tdma-based media access control
  (mac) protocol for mobile sensor networks.
\newblock In Tomasz Janowski and Hrushikesha Mohanty, editors, {\em ICDCIT},
  volume 4882 of {\em Lecture Notes in Computer Science}, pages 42--53.
  Springer, 2007.

\bibitem{Johnen09}
Colette Johnen and Le~Huy Nguyen.
\newblock Robust self-stabilizing weight-based clustering algorithm.
\newblock {\em Theor. Comput. Sci.}, 410(6-7):581--594, 2009.

\bibitem{KM06c}
Hirotsugu Kakugawa and Toshimitsu Masuzawa.
\newblock A self-stabilizing minimal dominating set algorithm with safe
  convergence.
\newblock In {\em 20th International Parallel and Distributed Processing
  Symposium (IPDPS 2006)}, 2006.

\bibitem{KP98}
S.~Kutten and D.~Peleg.
\newblock Fast distributed construction of small-dominating sets and
  applications.
\newblock {\em Journal of Algorithms}, 28(1):40--66, 1998.

\bibitem{L78}
L.~Lamport.
\newblock Time, clocks and the ordering of events in a distributed system.
\newblock {\em Communications of the ACM}, 21(7):558--565, 1978.

\bibitem{PB04}
L.~D. Penso and V.~C Barbosa.
\newblock A distributed algorithm to find $k$ -dominating sets.
\newblock {\em Discrete Applied Mathematics}, 141(1-3):243--253, 2004.

\bibitem{S90}
F.B. Schneider.
\newblock Impliementing fault tolerant services using the state machine
  approach: a tutorial.
\newblock {\em Computing Surveys}, 22(4):299--319, 2990.

\bibitem{S02}
I.~Stojmenovic.
\newblock {\em Handbook of Wireless Networks and Mobile Computings}.
\newblock John Wiley \& Sons, 2002.

\end{thebibliography}
\end{document}